\documentclass[conference]{IEEEtran}
\IEEEoverridecommandlockouts
\usepackage{cite}
\usepackage{amsmath,amssymb,amsfonts}
\usepackage{algorithmic}
\usepackage{graphicx}
\usepackage{textcomp}
\usepackage{xcolor}
\def\BibTeX{{\rm B\kern-.05em{\sc i\kern-.025em b}\kern-.08em
    T\kern-.1667em\lower.7ex\hbox{E}\kern-.125emX}}
\begin{document}

%
\title{Investigating team maturity in an \\agile automotive reorganization}

\author{
\IEEEauthorblockN{Lucas Gren}
\IEEEauthorblockA{\emph{Chalmers $|$ University of Gothenburg}\\Gothenburg, Sweden\\
Email: lucas.gren@cse.gu.se}
\and
\IEEEauthorblockN{Niclas Pettersson}
\IEEEauthorblockA{\emph{Volvo Cars}\\ Gothenburg, Sweden\\
Email: niclas.pettersson@volvocars.com}

}

\maketitle

\begin{abstract}
About seven years ago, Volvo Cars initiated a large-scale agile transformation. Midst this journey, a significant restructuring of the R\&D department took place. Our study aims to illuminate how team maturity levels are impacted during such comprehensive reorganizations. We collected data from 63 teams to comprehend the effects of organizational changes on these agile teams. Additionally, qualitative data was gathered to validate our findings and explore underlying reasons.
Contrary to what was expected, the reorganization did not significantly alter the distribution of team maturity. High turnover rates and frequent reorganizations were identified as key factors to why the less mature teams remained in the early stages of team development. Conversely, teams in the second category remained stable at a higher maturity stage, primarily because the teams themselves remained largely intact, with only management structures changing.
In conclusion, while reorganizations may hinder some teams’ development, others maintain stability at a higher level of maturity despite substantial managerial changes. 
\end{abstract}

\IEEEpeerreviewmaketitle

\section{Introduction}\label{sec:introduction}
An agile approach to projects has spread also to the automotive industry.  The possibility of changing what is being built faster has also become a competitive advantage when manufacturing cars. Quicker feedback loops connected to actual customer value is needed for many car components rather than being subject to the uncertainties of large up-front design and rigorous planning \cite{isleanagile}. 

Having systems where the requirements do not change across time is becoming more and more rare in the engineering context and especially in software engineering. Since the automotive industry has experienced an explosion of software components, large up-front designs tend to result in a lot of waste. Research in automotive software engineering focuses on testing and building software models \cite{haghighatkhah2017improving}. There are very few studies on organizational psychology in automotive software engineering in general, one notable exception being studies on ``human factors'' on the individual level (e.g.\ \cite{maro2017challenges}). We have not been able to find any studies on social psychology in general, or team maturity in particular, in automotive software development. 

There are studies in general software engineering, however, on self-organizing teams (e.g., \cite{moe2008}), different emerging roles in agile teams (e.g., \cite{hoda2012self}), and agile teams' adoption of agile practices (e.g., \cite{hodgson2013controlling}). In organizational psychology, the increased interest in self-organizing or self-managing groups, e.g.\ \cite{kaikkonen2018characteristics}, has resulted in an increased interest in team development and contemporary organizational contexts like, for example, lean implementations in health organizations \cite{ulhassan2014does}. In previous research, it has been found that agile teams can also benefit from emphasizing group development \cite{gren2020group}, which implies that it is an important component in the development of large software systems.


\subsection{The Volvo Cars Case}
In the last seven years, all of Volvo Cars' R\&D division has gone through an agile transformation, i.e.\ a ``big bang'' on a very large scale effecting around 750 teams.  To address the emerging technical challenges posed by Autonomous Drive, Electrification, and Digitalization in the automotive industry, Volvo Cars recognized the need for empowered and mature agile teams. These teams collaborated to devise innovative technical solutions, leveraging their collective creativity. From a psychological standpoint, mature teams play a pivotal role in unlocking agility and fostering innovation \cite{gren2020agile}. Volvo Cars aimed to cultivate such teams on a large scale to drive its technological advancements.

To address the reflexivity \cite{lenberg2023qualitative} in this study which is so influenced by the cultural context, we believe it is important to explain our context and philosophy of work at Volvo Cars. 
We believe that leadership has an essential role to play if more teams should reach higher maturity (team maturity is explained in Section 1.B). Leading towards self-organization is hard and new to many. We have also seen cases where teams are reluctant to take responsibility, which could be due to multiple reasons (individual or related to group norms\slash organizational culture). What is important to realize is that every single employee needs to change for the company to harness the benefits of a team-based and agile automotive company. In an agile world, responsibility exists together with a real mandate to succeed and fail. Hierarchies of expertise exist but with a purpose to support and not control. Very skilled and brave leaders exist that defend the new ways of collaborating, which means we can adjust unwanted behavior both below us in the hierarchy, but just as important, to peers on our own level and upward. Cooperation trumps competition. Added value trumps ownership. The best team outperforms the best individual. Offering help nurtures innovation which controlling inhibits. 

We summarized a list of 10 external items that teams need in order even have the possibility to mature further: (1)Thought-through reorganizations (teams in mind), (2) Focus on teams (not individuals), for example, Team rewards, (3) Management of team-promoting ecosystems, (4) Teams with 3-9 members, (5) Clear mandate (framing for\slash purpose of teams), (6) Dedicated support for team development (needed at every department), (7) Priority of team development in backlogs, (8) Leaders that support (and support in the right way), (9) Value team recommendations (management does not think for the teams), and (10) The right human and technical resources.

A large part of succeeding in having a team-based agile setup is to re-onboard the leadership in the new agile culture by training, but also changing the promotion\slash recruitment processes (i.e. to choose high trust over high performance, if one has to choose).

We observed a reluctance to focus on people and interactions over processes and tools in general and saw this as one of the largest forces against agility on a larger scale. Instead of focusing on processes and tools, we would like a focus on helping the understanding of where we think the toughest challenges are; the cultural transformation. 
As a part of the cultural transformation, at Volvo Cars we communicated the following manifesto as a guidance when employees were confused about how to act: (1) Internal cooperation over internal competition, (2) Added value over departmental ownership, (3) Strong teams over strong individuals, and (4) Offering help over exercising control.  That is, while there is value in the items on the right, we value the items on the left more.


\subsection{Introduction to Team Development}
Wheelan \cite{wheelandev} developed the Integrated Model of Group Development (the IMGD). It integrates earlier models of group development suggested by Tuckman \cite{tuckman1965developmental}, Bion \cite{bion}, and Bennis and Shepard \cite{bennis1956theory}, and among others. The IMGD describes how small groups move through four distinct stages and a termination stage \cite{wheelan}:

Stage 1---Dependency and inclusion: At this stage, members are dependent on the leader and concerned with issues around inclusion. The goal is to establish safety and sufficient structure for the group to start working. Members are hesitant to express divergent views.

Stage 2---Counter-dependency and fight: With a growing sense of safety, members start expressing different views. The group works on integrating these and freeing themselves from their dependence on the leader. Groups that have difficulties integrating their differences may resort to scapegoating the leader and\slash or individual members. The goal in Stage 2 is to solve task conflicts, avoid personal conflicts, and to clarify goals and work procedures.

Stage 3---Trust and structure: The integration of differences and reallocation of authority moves the group into Stage 3. When members find that different views are accepted, trust increases. Members realize that they are interdependent. The focus now is the balance between interdependence and individual autonomy and the continued clarification of goals, roles and work processes.

Stage 4---Work and productivity: As a result of the continued work on roles, goals and work processes, the group reaches the highest stage of development, Stage 4. Effectiveness, cohesion and work satisfaction are high. The group works as a team. Leader dependence is low.

Termination---Many groups don’t have a preset end date, but some do. The termination stage is an opportunity to look back and learn from experiences. 

Wheelan created the Group Development Questionnaire (the GDQ) \cite{wheelan}, which is a 60-item survey measuring the four development stages described above. 

\subsection{The Team Maturity Measurement Tool}
Between 2018 and 2021, Volvo Cars trained 27 in-house trainers who, in turn, trained more than 1,000 teams (750 active since some teams were terminated while others were formed) in a 3h introductory workshop. The workshop was based on Wheelan's theory of the Integrated Model of Group Development (IMGD). The final part of the training was to familiarize the team with the Team Health Check application created in-house at Volvo Cars implementing the a short version of the Group Development Questionnaire. The GDQ Short is a reduced version of the GDQ and comprises only 13 items ideal for faster assessments over time \cite{gren2020group}. Teams get a match in one of the four developmental stages if they are more than one standard deviation from the norm data mean value, which is from a large non-Volvo Cars dataset. For more details in the short team maturity measurement, see \cite{gren2020group}. It is important to note that team get matches in multiple stages which would categorize them as split. If team members think differently about the same team there are subgroups  \cite{hornsey2000assimilation} in the team resulting in this categorization. 

Getting the three-hour training was a prerequisite for using the tool. In doing so, we hoped they would feel empowered and the teams owned their own results, meaning that no one in the company could see their data but themselves.


In summary, the team maturity measurement implemented in the THC was in place both before and after the line organization decided to reorganize R\&D enabling an analysis of what happened to the teams' maturity.  We would expect a change in teamwork for the worse if the management structure changes and organizational chart is fundamentally redrawn, which was the case. We therefore formulated the following research question: 

RQ: What happens to the teams' maturity levels during a reorganization in a large scale agile automotive company?




\section{Method}\label{sec:methodology}
The reorganization took place on October 1, 2020.  The analysis was deemed too sensitive to publish for a couple of years, which is why it is first published in 2024. To investigate the effect of the reorganization, we compared all the team measurements from the teams who conducted one measurement six months before the change, and one six months after.  If a team had conducted more than one measurement before or after, we only used the most recent measurement. 

In total, 549 teams had conducted a measurement in the period of six months before or six months after, however, only 63 teams had conducted both. Hence, we only used 63 teams in our analysis. 
We could not influence who used the tool since the teams who had taken the training chose the frequency themselves.  
We assess the differences by a calculated distributions of percentages before and after the organizational change.





\paragraph{Procedure}
We sorted the dataset based on timestamps and anonymous team identifiers. The rationale behind looking at data six months before and six months after the reorganization was that teams are recommended to use the tool at least in every program increment of three months. However, not all teams do and we wanted to capture as many teams as possible while still keeping a quite current measurement from around the date of the reorganization.  We used RStudio and wrote all our scripts in R. All the R scripts and some example data is shared on Zenodo [link added upon acceptance].

We also presented the result to the Team Maturity Community of Practice (CoP) at Volvo Cars in October 2021 for an audience of around 10-20 employees (some listen in to parts of the Zoom call).  We presented the distributions before and after the reorganization and asked the audience what they thought might be the reasons for the distributions before and after. 

\paragraph{Participants}
The data from the team measurements came from a variety of teams at Volvo Cars' Research and Development organization.  The organization consisted of several departments like Software and Electronics, Commercial, Propulsion, Complete Vehicle, and Hardware to name a few. The total organization comprised of around 10,000 employees and 750 teams with team sizes from three to 15 members.  The subset used for this study were from all the departments.  Out of the selected 63 teams, 27 were pure software development teams, 23 were mixed with both hardware and software development within the same team, and 13 were pure hardware development. 


The focus group consisted of employees who were trainers of the Team Health Check tool and part of the Team Maturity Community of Practice (or the THC CoP).  The Team Maturity CoP took place online biweekly and was a forum for discussion on all things related to the development of agile teams at Volvo Cars. The trainers at the CoP were on all levels and most roles including Scrum Masters, Release Train Engineers, Line Managers, Change Leaders, and Developers. 

The data go beyond only software development, however, both the agile approach to projects and the Scaled Agile Framework stem from software engineering. In addition, most automotive development today has connection, often in the same team, to software engineering and in our sample 79\% had software developed in the team.

\section{Results}\label{sec:results}
This section presents both the qualitative and the quantitative results. 

\paragraph{Quantitative results}
Table~\ref{before} shows the data from six months before the reorganization of October 1, 2020 (with no multiple measurements, i.e. only the most recent measurement from each team in that period), and from six months after for the same set of teams.

\begin{table}\caption{Data from before and after the reorganization.}\label{before}
\centering
\begin{tabular}{cccccc}
\hline
Before: & Split &  Stage 1 & Stage 2 & Stage 3 & Stage 4 \\
&21\% &  24\%& 3\% &43\%&9\%\\
\hline
After: & Split &  Stage 1 & Stage 2 & Stage 3 & Stage 4 \\
&27\% &  22\% & 3\% & 35\% &13\%\\
\end{tabular}
\end{table}




The examination of percentages reveals a notable absence of discernible differences. Following a substantial reorganization, the distribution remains similar. This observation was surprising to us initially as our collective experience suggests that reorganizations typically regress teams in their developmental progress. 

\paragraph{Qualitative results}
During discussions within the THC CoP, participants offered insights into the factors influencing the observed outcomes. Firstly, the presence of high turnover and frequent reorganizations appeared to leave teams either fragmented or stranded in Stage 1. These teams found themselves in the initial phases of comprehending their roles and exhibiting lower levels of psychological safety. The persistence of traditional command-and-control leadership exacerbated this situation, as it inhibited the essential questioning and internal restructuring required for team effectiveness. 

An explanation for the substantial number of teams maintaining Stage 3 maturity levels within such a volatile environment lied in the fact that many teams remained structurally intact during the reorganization, while only the management changed. In numerous instances, new Team Managers (first level line managers), Scrum Masters (ScMs), or Product Owners were appointed, yet the composition of developers within these teams remained stable. Consequently, the internal collaboration within these teams retained a relatively mature level, even with the introduction of a new organizational structure featuring distinct roles.

\section{Discussion}\label{sec:discussion}

The distribution of teams six months before and after the reorganization were similar in this study. Based on previous research on team development, one would hypothesize a large reorganization to push many teams back in their development cycle \cite{wheelandev}. However, this implies a close-knit integration of management roles into teams. The discussion in the focus group, on the one hand, confirmed the hypothesis of teams remaining split or reset in Stage 1, however, the explanations to why so many teams remain in Stage 3 deserves some further reflection. 

The presence of stable and mature teams is undeniably beneficial.  However, there is a caveat to consider: if the connection between the management and these teams becomes so detached that it renders management impervious to changes in their structure, it may hinder optimal support for the teams. The line organization, with its defined responsibilities, has the potential to assist teams in addressing various impediments, which are often crucial for teams to advance to Stage 4. While Stage 3 teams excel in optimizing their internal processes, they may not realize their full potential in terms of seamless collaboration with other teams, as highlighted by previous research \cite{wheelandev}. Stage 3 serves as a significant stepping stone, but the true aspiration lies in Stage 4, where teams foster innovation and consistently deliver high-quality results in partnership with their stakeholders.

Our capacity to elevate more teams to a state of maturity is intrinsically tied to the degree to which future reorganizations preserve the integrity of these teams. It is crucial to recognize that any modification in the composition of a team setup directly impacts the team's maturity level. Consequently, it is advisable to prioritize team-level changes over individual-level alterations during reorganizations, with the aim of promoting team maturity. This approach was notably emphasized in the case exemplified within this study, where team-level modifications were prioritized to a significant extent.








\section{Threats to validity}\label{sec:limits}

The GDQS on which the THC is built, were subjected to a diversity of validation studies \cite{gren2020group}.  We therefore assess the quantitative measurement as quite stable and consistent over time.  The qualitative data on the other hand would be quite hard to replicate if not asking exactly the same people again. This is intrinsic to qualitative data, which is, however, much richer in finding some explanations for the found result. 

There is a large risk in trying to measure something as complex as group dynamics (or group development specifically) with only 13 items. Such a complex construct surely comprises much more than what is captured. However, the constructs in group development are well-researched (e.g.,  \cite{tuckman}) and these 13 items are believed to capture some aspects of the theory of group development. 

We mostly use our qualitative data to draw conclusions of the patterns seen in the distributions before and after the reorganization that took place.  However, these conclusion are believed to be reasonably accurate for the Volvo Cars case and need further investigation outside of the company. 

There is a possible selection bias since we do not know why some teams start to measure and others do not.  This would induce confounding factors that we are not aware of. Another threat to internal validity is that the GDQS used in the THC actually does not capture team development in the way that is believed. 

We have very scarce data on potential explanation for our result and broader generalizations should be done with care. Maybe other auto-manufacturers with a team-based R\&D organization would have similar results, but this remains to be explored.

\section{Conclusion and Future Work}\label{sec:limits}
This study examined how agile team maturity fared during a major company reorganization. Interestingly, the distribution of team maturity remain stable throughout the change. To further explore these findings, replicating this study at other companies is recommended. The study highlights the critical role that social psychological team factors play in the success of agile teams, especially for the future of car development.



\bibliographystyle{IEEEtran}

\bibliography{ref}

\end{document}